\documentstyle[11pt]{article}
\normalbaselines
\begin{document}
\bibliographystyle{unsrt}

\def\bea*{\begin{eqnarray*}}
\def\eea*{\end{eqnarray*}}
\def\ba{\begin{array}}
\def\ea{\end{array}}
\count1=1
\def\be{\ifnum \count1=0 $$ \else \begin{equation}\fi}
\def\ee{\ifnum\count1=0 $$ \else \end{equation}\fi}
\def\ele(#1){\ifnum\count1=0 \eqno({\bf #1}) $$ \else \label{#1}\end{equation}\fi}
\def\req(#1){\ifnum\count1=0 {\bf #1}\else \ref{#1}\fi}
\def\bea(#1){\ifnum \count1=0   $$ \begin{array}{#1}
\else \begin{equation} \begin{array}{#1} \fi}
\def\eea{\ifnum \count1=0 \end{array} $$
\else  \end{array}\end{equation}\fi}
\def\elea(#1){\ifnum \count1=0 \end{array}\label{#1}\eqno({\bf #1}) $$
\else\end{array}\label{#1}\end{equation}\fi}
\def\cit(#1){
\ifnum\count1=0 {\bf #1} \cite{#1} \else 
\cite{#1}\fi}
\def\bibit(#1){\ifnum\count1=0 \bibitem{#1} [#1    ] \else \bibitem{#1}\fi}
\def\ds{\displaystyle}
\def\hb{\hfill\break}
\def\comment#1{\hb {***** {\em #1} *****}\hb }

\newtheorem{theorem}{Theorem}
\newtheorem{proposition}{Proposition}
\newtheorem{lemma}{Lemma}

\newcommand{\TZ}{\hbox{\bf T}}
\newcommand{\MZ}{\hbox{\bf M}}
\newcommand{\ZZ}{\hbox{\bf Z}}
\newcommand{\NZ}{\hbox{\bf N}}
\newcommand{\RZ}{\hbox{\bf R}}
\newcommand{\CZ}{\,\hbox{\bf C}}
\newcommand{\PZ}{\hbox{\bf P}}
\newcommand{\QZ}{\hbox{\bf Q}}
\newcommand{\HZ}{\hbox{\bf H}}
\newcommand{\EZ}{\hbox{\bf E}}
\newcommand{\GZ}{\,\hbox{\bf G}}
\newcommand{\UZ}{\hbox{\bf U}}
\font\germ=eufm10
\def\goth#1{\hbox{\germ #1}}
\newtheorem{pf}{Proof}
\renewcommand{\thepf}{}
\vbox{\vspace{38mm}}
\begin{center}
{\LARGE \bf Structure of Certain Chebyshev-type Polynomials in Onsager's Algebra Representation \footnote{Talk presented in "AMS-SIAM Special Section on Integrable Systems and Special Functions", Atlanta, GA, USA, January 2005.}} \\[5mm]
 Shi-shyr Roan \\{\it Institute of Mathematics , 
Academia Sinica \\  Taipei , Taiwan \\ (e-mail:
maroan@gate.sinica.edu.tw)} \\[5mm]
\end{center}

\begin{abstract} 
In this report, we present a systematic account of mathematical structures of certain special polynomials arisen from the energy study of the superintegrable $N$-state chiral Potts model with a finite number of sizes. The polynomials of low-lying sectors are represented in two different forms, one of which is directly related to the energy description of superintegrable chiral Potts $\ZZ_N$-spin chain via the representation theory of Onsager's algebra. Both two types of polynomials satisfy some $(N+1)$-term recurrence relations, and  $N$th order differential equations; polynomials of one kind reveal certain Chebyshev-like properties. Here we provide a rigorous mathematical argument for cases $N=2, 3$, and further raise some mathematical conjectures on those special polynomials for a general $N$. 
\end{abstract}
 \par \noindent
1999 PACS: 03.65.Fd , 02.30.Hq , 75.10.Jm \par \noindent
$Keywords:$ Onsager's algebra, Chiral Potts $\ZZ_N$-spin chain , Chebyshev-type polynomials

\vfill
\eject
\section{Introduction}
This article contains a sound mathematical treatment of certain  special polynomials appeared in the eigenvalue problem of superintegrable chiral Potts $\ZZ_N$-spin chain of a finite size through Onsager's algebra representation. Throughout this paper, $N$ will always denote an integer $\geq 2$, $\omega= e^{\frac{2\pi {\rm
i}}{N}}$, and $\CZ^N $ is the vector space consisting of $N$-cyclic vectors with the basis 
$\{ | n \rangle \}_{ n \in \ZZ_N}$. Denote by $X, Z$
the operators of $\CZ^N $ satisfying the Weyl  
relation and $N$th power identity property: $XZ= \omega^{-1}ZX$, $X^N=Z^N=1$. One can represent $X, Z$ in matrix forms by  $X |n \rangle = | n +1 \rangle$, $ Z |n \rangle = \omega^n |n \rangle $ for $n \in \ZZ_N$. 
The Hamiltonian of the superintegrable chiral Potts  
$\ZZ_N$-spin chain is the operator acting on $\stackrel{L}{\otimes}\CZ^N$ defined by
\be
H(k') = - \sum_{l=1}^L \sum_{n=1}^{N-1}
\frac{2}{1-\omega^{-n}}( X_l^n +
k'Z_l^nZ_{l+1}^{N-n}) 
\ele(SCPM)
where $k'$ is a real parameter, and $X_l, Z_l$ are operators $X$, $Z$ acting on $\CZ^N $ at the site $l$ with periodic boundary condition: $Z_{L+1}= Z_1$. For $N=2$, this is the Ising quantum chain. The Hamiltonian $H(k')$ commutes with the spin-shift operator $\bigotimes_{l=1}^L X_l$, hence carries the
$\ZZ_N$-charge, denoted by $Q = 0 , \ldots, N-1$, in this paper. 
One can split $H(k')$ into a sum of Dolan-Grady pair of operators with the coupling parameter $k'$, by which the representation theory of Onsager's algebra can be employed for  the study of $H(k')$-eigenvalue problem \cite{GR}. However for its spectrum calculation, the available method relies on the $N$-state chiral Potts model, a two-dimensional solvable statistical lattice model, where $H(k')$ is derived as the derivative of transfer matrix at the superintegrable rapidity point (see, e.g. \cite{AMP, B} and references therein). For the superintegrable chiral Potts model with a finite size, the eigenvalues of transfer matrices are given by certain polynomial solutions of a spin-$\frac{N-1}{2}$ XXZ-model-like Bethe equation \cite{AMP, B, B93, B94, DKM}. The low-lying (no Bethe-excitation) sector polynomials of chiral Potts transfer matrices for the ground-state eigenvalue were explicitly constructed in \cite{AMP, B, B94}, and will be called the Baxter-Albertini-McCoy-Perk (BAMP) polynomials in this article. By converting the chiral Potts  eigen-polynomials to another ones through a simple change of variables, one can determine 
the eigenvalues of the Hamiltonian (\req(SCPM)). In particular, the BAMP polynomials are transfered to another type of polynomials, which are more convenient to express the spectra of (\req(SCPM)). The resulting polynomials in the Ising case turn out to be Chebyshev polynomials. For $N \geq 3$, these polynomials have shown many properties reminiscent of, but not fully in agreement with orthogonal polynomials, hence will be called the (generalized) Chebyshev-type polynomials in this paper. In \cite{G02, GRo}, we have made empirical calculations on these polynomials  for small $N$ by investigating the recurrence relations, differential equations etc., in hope to explore the zero distributions of those Chebyshev-type polynomials. The computations have shown a remarkable feature of "partial orthogonality" in terms of Jacobi weights, leading to a certain conjecture  about those Chebyshev-type polynomials (for the details, see \cite{G02}). Through explicit calculations made in the study of these polynomials in the context of superintegrable chiral Potts models, we discovered certain simple mathematical structures in the process. It is the aim of this report to present a mathematical  account on the structures of the BAMP and Chebyshev-type polynomials for all $N$ regarding to the recurrence relations and differential equations, based upon the computational results found in \cite{G02, GRo}. The clarified mathematical presentation of the subject in this article, we hope, could stimulate some interest in these special polynomials with a sound physical interest, however unexplored in classical literatures ( e.g. \cite{Chi, S}) to the best of our knowledge.

The paper is organized as follows. In Sec. 2, we review some basic facts on Onsager's algebra and its connection with the Hamiltonian (\req(SCPM)). In Sec. 3, we define the BAMP polynomials appeared in the spectra of $N$-state chiral Potts models, and discuss  their common mathematical structures for all $N$. In the Ising case, the relation between  ground state energy and all other eigenvalues are given. For a general $N$, the recurrence relation of BAMP  polynomials by varying the size $L$ modular $N$ is systematically constructed, and certain qualitative properties related to the recurrence relation are discussed.  The problem of hypergeometry-like higher order differential equations governing the BAMP polynomials are raised in Sec. 4 with illustrative discussions on the case $N=2,3$.  In Sect. 5, we define the Chebyshev-type polynomials, and discuss their recurrence relations. In the Ising case, we provide a detailed account on its relation with  Chebyshev polynomials.   
In Sect. 6, we discuss the problem of differential equations for Chebyshev-type polynomials, and present a rigorous mathematical proof of 
solutions for those "$\ZZ_3$-Chebyshev-type" polynomials.

\section{Onsager's Algebra}
In the seminal paper of Onsager on the free energy solution of the two-dimensional Ising model \cite{O}, there appeared an infinite-dimensional Lie algebra, called Onsager's algebra, generated by a basis $\{ A_m, G_l \}_{ m \in \ZZ , l \in \NZ }$ with the following commutation relations:
$$
[ A_m , A_n ] = 4 G_{m-n} \ , \ \ \ [ A_m , G_l ] = 2 A_{m-l} - 2 A_{m+l} \ , \ \ \ [G_m , G_l ] = 0 \ ,
$$
where $G_{-l} := - G_l $ and $G_0 : = 0$. The elements $A_0 , A_1$ satisfy the Dolan-Grady relation \cite{DG}:
$$
[ A_1 , [ A_1 , [A_1 , A_0 ]]] = 16 [A_1 , A_0 ] \ , \ \ \ [ A_0 , [ A_0 , [A_0 , A_1 ]]] = 16 [A_0 , A_1 ] \ ,
$$
and Onsager's algebra is the Lie algebra generated by this Dolan-Grady pair $\{ A_0, A_1 \}$ \cite{D91, R91}.  Inspired by results in \cite{D90} on the (finite-dimensional) unitary representations of Onsager's algebra, a realization of Onsager's algebra as the Lie-subalgebra of the loop algebra ${\goth sl}_2 [z, z^{-1}]$ fixed by a standard involution was found in \cite{R91} through the identification:
$$
A_m = 2 z^m e^+ + 2 z^{-m} e^- \ , \ \ G_m = (z^m - z^{-m}) h \ , \ \ \ \ m \in \ZZ ,
$$
where $e^\pm , h$ are generators of ${\goth sl}_2 $ with $[e^+, e^-]=h$, $[h, e^{\pm} ] = \pm 2 e^\pm $. By which, a thorough mathematical study of  Onsager's algebra was carried out in \cite{DR} by using techniques in ideal theory of polynomial algebras, and the discussions further enrich our understanding the various aspects of the mathematical structures of  Onsager's algebra. It is known that all finite-dimensional irreducible representations of the loop algebra ${\goth sl}_2 [z, z^{-1}]$ are given by tensoring a finite number of irreducible ${\goth sl}_2$-representations through the evaluation  of $z$ at distinct non-zero complex values $a_j$s. The Hermitian irreducible representations of Onsager's algebra are obtained by factoring Onsager's algebra through the ${\goth sl}_2 [z, z^{-1}]$-representations, with further constraints on the evaluated values $a_j$'s: $| a_j | = 1, a_j \neq \pm 1$ and $a_j \neq a_k^{\pm 1}$ for $j \neq k$. By which,  
$A_m = 2 \sum_{j=1}^n ( a_j^m e^+_j + a_j^{-m} e^-_j)$, $ G_m = \sum_{j=1}^n (a_j^m - a_j^{-m}) h_j$. 
For the Hamiltonian (\req(SCPM)), we write $H(k')$ as a sum of Dolan-Grady pair $\{ A_0, A_1 \}$ with the parameter $k'$: $H(k') = - \frac{N}{2} ( A_0 + k' A_1 )$. By \cite{B}, all the ${\goth sl}_2$-irreducible representations involved in the spectra of $H(k')$ are of spin-$\frac{1}{2}$, hence their eigenvalues have the Ising-like expression:
$$
\alpha + \beta k'  + N \sum_{j=1}^n \pm \sqrt{1 - 2 k' \cos(\theta_j) + k'^2} \ , \ \ \ \ \cos(\theta_j) = \frac{-1}{2}( a_j + a_j^{-1}) \ , \ \ \ \alpha , \beta \in \RZ \ .
$$
The number and location of $c_j$'s, depending on the chain-size $L$ and $\ZZ_N$-charge $Q$, are the ingredients for the eigenvalues of $H(k')$. For the rest of this paper, we shall mainly discuss issues related to the ground-state eigenvalue. 

\section{Recurrence
Relation of Baxter-Albertini-McCoy-Perk (BAMP) 
Polynomials}
We start with some simple elementary algebra for later
use. For a given $N$, the symbol $\underline{m}$ for $m \in \ZZ$ will always denote 
the integer, $ 0 \leq \underline{m} \leq N-1$, with 
$\underline{m} \equiv m \pmod{N}$. Let $\CZ[t], \CZ[s]$ be the polynomial algebras in  variables $t, s$ respectively with the relation $s: =
t^N$. Then any element $f (t)$
in $\CZ[t]$ can be uniquely expressed by $
f ( t ) = \sum_{j=0}^{N-1} t^j F_j (s)$ with $
F_j (s) \in \CZ [s]$. 
The multiplication by $\sum_{j-0}^{N-1} t^j$ on
$\CZ[t]$, $
f (t) \mapsto g(t) = ( \sum_{j=0}^{N-1} t^j ) f
(t) $, gives rise to an endomorphism of
$\CZ[t]$, which can be represented by the following matrix-form (by considering $\CZ[t]$ as a $\CZ[s]$-module ) with $
g ( t ) = \sum_{j=0}^{N-1} t^j G_j (s)$,
\be
\left( \begin{array}{c}
       G_0(s)\\
       G_1(s) \\
       \vdots \\ 
      G_{N-1}(s)\\  
\end{array} \right) = R \left( \begin{array}{c}
       F_0(s)\\
       F_1(s) \\ 
      \vdots \\ 
       F_{N-1}(s)\\  
\end{array} \right) \ ; \ \ R: = \left(
\begin{array}{cccc}
       1  & s & \cdots   & s \\
        1 & 1 &  \ddots   &\vdots \\  
      \vdots  & \ddots  & \ddots & s
\\ 
 1 & \cdots  &  1& 1 \\  
\end{array} \right) \ .
\ele(R)
Denote by $A(x)$ the characteristic polynomial
of the matrix $R^N$:
\be
A(x) = 
{\rm det}( x- R^N) = x^N  + \ldots
+ A_k(s) x^{N-k} + \ldots + A_N(s) \ .
\ele(chRN)
By the
Hamilton-Cayley theorem, we have 
$$
A(R^N) =(R^N)^N + \ldots
+ A_k(s) (R^N)^{N-k} + \ldots + A_N(s) = 0 \ .
$$

For an arbitrarily given initial polynomial $b(t)$,
we consider a sequence of $s$-polynomials $F_{l, j} (s)$ for $l
\in \ZZ_{ \geq 0 }$ and $ 0 \leq j \leq N-1$,  defined by the
relation: 
\be
(\sum_{i=0}^{N-1} t^i)^{Nl} b(t) =
\sum_{j=0}^{N-1} t^j F_{l, j} (s) \ .
\ele(Flj)
Then $A(R^N) \bigg( F_{l,0}(s) , F_{l,1}(s) , 
       \ldots,  
       F_{l, N-1}(s) \bigg)^t = 0 $ for $ l \in \ZZ_{ \geq 0} $, 
equivalently, the following 
recurrence relation holds for $s$-polynomials $F_{l, j}(s)$ with a fixed $j$,
\be
F_{l+N, j} (s) + \ldots
+ A_k (s) F_{l+N-k, j} (s) + \ldots + A_N (s)
F_{l, j}(s) = 0 \ , \ \ l \geq 0 \ .
\ele(recF)
Note that the same $(N+1)$-recurrence relation
for $F_{*, j}(s)$'s in above holds for any  given initial
polynomial $b(t)$. 
For $N=2, 3, 4$, the explicit forms of (\req(recF)) are given by
\bea(ll)
N=2, & F_{l+2,j} -2(s+1)F_{l+1,j}+(s-1)^2F_{l,j} = 0 , \\
 N=3, & F_{l+3, j}
-3(s^2+7s+1)F_{l+2, j}+3(s-1)^4 F_{l+1, j}-
(s-1)^6 F_{l, j} = 0 , \\
N=4, &  F_{l+4, j} -4(s+1)(s^2+30s+1)
F_{l+3, j} +2(3s^2-62s+3)(s-1)^4
F_{l+2, j} \\ &
-4(s+1)(s-1)^8F_{l+1, j}+
(s-1)^{12} F_{l, j} = 0 
\elea(F234)
For a general $N$, one can determine the values of $A_k(1)$ for all $k$. Indeed, by setting $s=1$ for the matrix $R$ in (\req(R)), one can extend the vector $(1, \ldots, 1)^t$ to a basis of $\CZ^N$ so that $R_{s=1}$ is expressed by ${\rm dia} [ N, 0 , \ldots, 0]$. Hereafter, we use ${\rm
dia} [\alpha_0, \cdots, \alpha_{N-1}]$ to denote the diagonal
matrix with entries $\alpha_j$. Hence the characteristic polynomial (\req(chRN)) of $R_{s=1}^N$ is equal to $(x -N^N)x^{N-1}$, which implies
\be
A_1 (1 ) = - N^N \ , \ \ \ \ A_k (1) = 0 \ \ {\rm for} \ \ \ k \geq 2 \ .
\ele(RNs1) 

For the eigenvalue problem of the transfer matrix of size $L$ in the superintegrable $N$-state chiral Potts model, the eigenvalue of the ground state corresponds to the initial polynomial
$b(t)$ being the constant function 1. In general, the eigenvalues of the
superintegrable chiral Potts model are solved by the method of Bethe equations in
\cite{B94} as follows. Let $f(t ;
\vec{v})$ be the $t$-polynomial of degree
$m_p$ with the parameter $\vec{v}=(v_1,
\ldots, v_{m_p})$ defined by $
f(t ; \vec{v}) = \prod_{j=1}^{m_p} ( 1 + \omega v_j t )$. Associated to $f(t ; \vec{v})$, we consider the following function $p ( t ; \vec{v} )$: 
\be
p ( t ; \vec{v} ) =  \frac{\omega^{-P_b}}{N}
\sum_{j=0}^{N-1}
\frac{(1-t^N)^L(\omega^jt)^{-P_a-P_b} }{(1-
\omega^jt)^L f(\omega^j t; \vec{v}) f(
\omega^{j+1} t ; \vec{v} )} \ ,
\ele(ptv)
where $P_a, P_b$ are integers 
satisfying  
$0 \leq P_a+P_b \leq N-1$, $P_b-P_a \equiv
Q+L \pmod{N}$, with $Q$ the
$\ZZ_N$-charge as before. For an
arbitrary $\vec{v}$,
$p ( t ; \vec{v} )$ is a rational function of $t$. The necessary and sufficient condition
for $p ( t ; \vec{v} )$ to be a $t$-polynomial is the following Bethe constraints of
$\vec{v}$,
\be
( \frac{v_j + \omega^{-1}}{v_j+ \omega^{-2}} )^L
= - \omega^{-P_a-P_b} \prod_{l=1}^{m_p}
\frac{v_j - \omega^{-1}v_l}{v_j - \omega v_l} \ ,
\ \ \ j = 1, \ldots, m_p \ .
\ele(CPMBe)
Note that up to a phase factor the above equation is the
Bethe equations for the spin-$\frac{N-1}{2}$, anisotropy $\gamma = \frac{\pi}{N} 
$ XXZ chain with $L$-size and the periodic boundary condition \cite{DKM}. For a Bethe-solution $\vec{v}$
of (\req(CPMBe)), $p ( t ; \vec{v} )$
is a $t$-polynomial, invariant under
the transformation: $t
\mapsto \omega t$, hence it can be written as a polynomial of
$s \ (: = t^N)$,
$$
p( t ; \vec{v}) = P (s ; \vec{v}) \ . 
$$  
Furthermore, the integers $P_a, P_b$ are chosen so that 
$P (0 ; \vec{v}) \neq 0$. The $s$-degree of the polynomial $
P(s; \vec{v})$  is given by $
{\rm deg} \ P (s; \vec{v}) = [\frac{(N-1)L
-P_a-P_b-2m_p}{N}]$ 
with all its roots real, which determine the
spectrum of (\req(SCPM)). Here $[r]$ denotes the integral part of a real number $r$. The eigenvalues of the ground state and one-excited
state are the polynomials for $m_p = 0$ and $1$ 
respectively.

For the Ising case, i.e. $N=2$, we have
$P_a+P_b =0, 1$. With
$f (t,\vec{v}) = 1$ in  (\req(ptv)), one has
$$
P^{(L)}_{P_a+P_b}(s=t^2)
= \frac{(-1)^{P_b}t^{-P_a-P_b}}{2}
\bigg( (1+t)^L + (-1)^{P_a+P_b} (1-t)^L \bigg) = \pm
\prod_{j=1}^{[\frac{L}{2}] } (s - s_j ) \ , 
$$
whose zeros are known in \cite{Bax}:
$s_j = - \tan^2\{ \frac{\pi}{L}(j-
\frac{1-P_a-P_b}{2}) \}$. 
By $\omega =-1$, the right hand side of Bethe equations
(\req(CPMBe)) is always equal to  $-\omega^{-P_a-P_b}$. 
Hence the solution of (\req(CPMBe)) is given by a collection of non-zero $v_j$'s such that  $v_j^{-1}$ are roots of the above
$t$-polynomial $P^{(L)}_{P_a+P_b}$. Hence the
polynomial $p(t, \vec{v})$ corresponding to a Bethe-solution of (\req(CPMBe)) has the form (up to a scale multiple): $
p(t, \vec{v}) = P(s= t^2, \vec{v}) \ \sim \ \pm
( \prod_{j_1< \ldots < j_m} s_j )
\prod_{ j \neq j_i , j = 1}^{ [\frac{L}{2}] } (s - s_j )$.

For $N \geq 3$, the eigen-polynomial for the ground state of $L$-size transfer matrix in superintegrable chiral
Potts model is determined by (\req(ptv)) by setting $P_b=0$,
and $ f(t; \vec{v})=1$ \cite{AMP, B}, 
\be
P^{(L)}_Q(s) = 
 \frac{t^{-a_{L, Q}}}{N}
\sum_{j=0}^{N-1}
\frac{\omega^{-ja_{L,Q}}(1-t^N)^L }{(1-
\omega^jt)^L } , \ \ \ a_{L, Q}: = \underline{-L-Q} \ .
\ele(PLQ)
Note that $a_{L, Q}= a_{L', Q}$ if $ L \equiv L' \pmod{N}$. The $P^{(L)}_Q(s)$ is a $s$-polynomial of degree 
$$
b_{L,Q}  = [\frac{(N-1)L-Q}{N}] . 
$$
We shall call $P^{(L)}_Q(s) $ the
Baxter-Albertini-McCoy-Perk (BAMP) polynomial.
For a given $L$,  the polynomials $P^{(L)}_Q(s)$, $0 \leq Q \leq N-1$, are characterized by following relation:
\begin{eqnarray}
 (\frac{t^N-1}{t-1})^L =
\sum_{Q=0}^{N-1} t^{a_{L,Q}} P^{(L)}_Q(s) \ , \ \ 
 a_{L, Q}= \underline{-L-Q} \ \ ( = (N-1)L-Q-Nb_{L,Q}) \ .
\label{Pdef}
\end{eqnarray}
By which $P^{(L)}_Q (s)$ are polynomials with positive integer coefficients. 
Furthermore, one has the following reciprocal property and
recurrence relations among BAMP polynomials. 
\begin{theorem} \label{thm:BAMP} 
(i) $P^{(L)}_Q(s)$ and $P^{(L)}_{Q'}(s)$ are  reciprocal to 
each other  if and only if 
$L+Q+Q' \equiv 0 \pmod{N} $.

(ii) Let $A_k(s)$ be the polynomials defined by $(\req(chRN))$. The {\rm BAMP}
polynomials $P^{(L)}_Q$ satisfy the recurrence relation by varying $L$ for the same $Q$,
\be
P^{(L+N^2)}_Q (s)+ A_1 (s) P^{(L+(N-1)N)}_Q(s)+ \ldots
+ A_k (s) P^{(L+(N-k)N)}_Q(s)+ \ldots + A_N (s)
P^{(L)}_Q(s) = 0 \ .
\ele(recP)

(iii) The polynomials $A_k(s)$ are characterized by the property $(\req(recP))$  for all $L$ and $Q$.  For $1 \leq k \leq N$,
$A_k(s)$ is a reciprocal polynomial of degree $ k(N-1)$ with the leading coefficient and $A_k(0)$ equal to $(-1)^k{N\choose k} $. Furthermore, $
A_1(s) = -N P^{(N)}_0(s) $, $ A_N(s) =
(-1)^N(s-1)^{N(N-1)} $, and 
for $k \geq 2$, one has $A_k (s) = (s-1)^{2 l_k} \overline{A}_k (s) $ with $\overline{A}_k (1) \neq 0$ for some positive integer $l_k$. 
\end{theorem}
{\it Proof.} 
By replacing $t$ by $\frac{1}{t}$
in (\req(Pdef)), one obtains $
(\frac{t^N-1}{t-1})^L = \sum_{Q=0}^{N-1}
t^Q s^{b_{L,Q}}P^{(L)}_Q(\frac{1}{s})$. Hence
the relation
$P^{(L)}_{Q'}(s)=
s^{b_{L,Q}}P^{(L)}_Q(\frac{1}{s})$ holds if and
only if  $L+Q+Q' \equiv 0 \pmod{N} $, then follows $(i)$. For $0 \leq m \leq N-1$, we consider the BAMP polynomials $P^{(L)}_Q$ with $L
\equiv m \pmod{N}$. 
Set $b(t) = (\sum_{i=0}^{N-1}t^i)^m$ in (\req(Flj)), then
$F_{l,j}(s) = P^{(L)}_Q(s) $ with $L = Nl+m$ and
$j = a_{L, Q}$. Hence 
(\req(recP)) follows from (\req(recF)), so we obtain $(ii)$.

Suppose that $A_k(s)$, $1 \leq k \leq N$, are polynomials such that the relation
(\req(recF)) holds for $P^{(L)}_Q (s)$ for all $L$ and $Q$. By (\ref{Pdef}) and the definition of the matrix $R$ in (\req(R)), one concludes that 
$R^{N(N-k)} + \sum_{k=1}^N A_k (s) (R^N)^k = 0$, hence $A_k(s)$'s are defined by
the relation (\req(chRN)). 
As the matrix $R^N$ corresponds to the multiplication of $(\frac{t^N-1}{t-1})^{N}$ on $\CZ[t] (= \sum_{j=0}^{N-1} t^j \CZ[s] )$, and $
(\frac{t^N-1}{t-1})^{N} = P^{(N)}_0(s) + \sum_{Q=1}^{N-1} t^{N-Q} P^{(N)}_Q(s)$, one has 
the matrix-form expression of $R^N$: 
$$
R^N = \left(
\begin{array}{ccccc}
       P^{(N)}_0  & s P^{(N)}_{1} &sP^{(N)}_2& \cdots  & sP^{(N)}_{N-1} \\
    P^{(N)}_{N-1} & P^{(N)}_0 & sP^{(N)}_1& \cdots   &\vdots \\ 
        \vdots & P^{(N)}_{N-1}&P^{(N)}_0 & \ddots   &\vdots \\ 
      \vdots  & \ddots & \ddots & \ddots & sP^{(N)}_1
\\ 
 P^{(N)}_{1} & P^{(N)}_2 & \cdots &  P^{(N)}_{N-1}& P^{(N)}_0 \\  
\end{array} \right) \ ,
$$
whose trace gives $A_1(s) = -N P^{(N)}_0(s)$. By ${\rm det} R = 
(1-s)^{N-1}$, one obtains $
A_N(s) = (-1)^N {\rm det} \ R^N$ =$ (-1)^N
(s-1)^{N(N-1)}$. 
It is easy to see that $P^{(N)}_0(s)$ is a monic polynomial of degree $(N-1)$ with $P^{(N)}_0(0)=1$, and $P^{(N)}_Q(s)$ for $ Q \neq 0$ are polynomials of degree $(N-2)$.  Therefore the  $(i,j)$th entry of $R^N$ is a polynomial of degree  $N-1$ or $N-2$, according to $i \leq j$ or $i>j$ respectively, which implies ${\rm deg} A_k(s) \leq k (N-1)$. 
Setting $s=0$ in (\req(chRN)), one obtains $(x-1)^N = x^N + \sum_{k=1}^N A_k(0) x^N$ hence $A_k(0) = (-1)^k{N\choose k}$.  
Substituting  $s$ by $s^{-1}$ in (\req(recP)), then multiplying the factor $s^{b_{L+N^2, Q}}$, by $(i)$ we obtain the following relation for $L$ and $Q'$ with $L+Q+Q' \equiv 0 \pmod{N}$:
$$
P^{(L+N^2)}_{Q'} (s)+ \ldots
+ s^{k(N-1)}A_k (\frac{1}{s}) P^{(L+(N-k)N)}_{Q'}(s)+ \ldots + s^{N(N-1)}A_N (\frac{1}{s})
P^{(L)}_{Q'}(s) = 0 \ .
$$
As the above relation of BAMP polynomials holds for all $L$ and $Q'$, by the characterization of $A_k(s)$ in $(ii)$, one obtains $
s^{k(N-1)}A_k (\frac{1}{s}) = A_k (s)$ for $ 1 \leq k \leq N $. 
Therefore $A_k(s)$ is a reciprocal polynomial of degree $k(N-1)$ with the leading term equal to $A_k(0) = (-1)^k{N\choose k}$. As the roots of a reciprocal polynomial consists of  pairs $\{ s_i, s_i^{-1} \}$ for $s_i \neq \pm 1$, together with certain possible multiple-roots at $s = \pm 1$, the conclusion for  $A_k (s)\ , k \geq 2,$ in $(iii)$ follows from (\req(RNs1)) and the reciprocal property of $A_k(s)$.
$\Box$ \par \vspace{.1in} \noindent
{\bf Remark} Since the coefficients of $P^{(L)}_Q(s)$ are all positive integers, its real roots must be negative. By the relation between $P^{(L)}_Q(s)$ and the eigenvalues of the Hermitian operator (\req(SCPM)) (which will be explained in the next section), all the roots of $P^{(L)}_Q(s)$ must be real numbers through the Onsager's algebra representation theory. However, a mathematical argument for this real-root property of $P^{(L)}_Q(s)$ purely from the polynomial-algebra viewpoint has not yet be found.
$\Box$ \par \vspace{.1in} \noindent

By (\req(F234)), we list the
recurrence relation of BAMP polynomials for
$N=2, 3, 4$ for later use:
\bea(ll)
N=2, & P_Q^{(L+4)} -2(s+1)P_Q^{(L+2)}+ 
(s-1)^2P_Q^{(L)} = 0 , \\
 N=3, & P^{(L+9)}_Q
-3(s^2+7s+1)P^{(L+6)}_Q+3(s-1)^4P^{(L+3)}_Q-
(s-1)^6P^{(L)}_Q = 0 , \\
N=4, &  P^{(L+16)}_Q -4(s+1)(s^2+30s+1)
P^{(L+12)}_Q+2(3s^2-62s+3)(s-1)^4
P^{(L+8)}_Q \\ &
-4(s+1)(s-1)^8P^{(L+4)}_Q +
(s-1)^{12} P^{(L)}_Q = 0 \ . 
\elea(P234)
For a general $N$,  there is a general expression of $A_k (s)$ as follows.
\begin{proposition} \label{prop:AP} Denote $\lambda_{\ell} = \sum_{Q=0}^{N-1} \omega^{Q \ell} s^{\frac{N-Q}{N}}$ for $0 \leq \ell \leq N-1$. Then the polynomials $A_k (s)$ for $1 \leq k \leq N$ in $(\req(recP))$ are expressed by
$$
A_k (s) = (-1)^k \sum_{\ell_1< \ell_2 < \ldots < \ell_k } \lambda_{\ell_1}^N \cdots \lambda_{\ell_k}^N .
$$
\end{proposition}
{\it Proof.} With $R$ in
(\req(R)) and  $
D:= {\rm dia} [s, s^{\frac{N-1}{N}} , \ldots , s^{\frac{1}{N}} ]$, $D^{-1} R D$ is the $N \times N$ Toeplitz (cyclic) matrix $\bigg( s^{(\underline{i-j})/N}\bigg)_{0 \leq i, j \leq N-1}$ with $s^{(\underline{i-j})/N}$ as the $(i, j)$th entry.  Hence  $\lambda_{\ell}$ is the eigenvalue of  $\bigg(s^{(\underline{i-j})/N}\bigg)_{i,j}$ with the eigenvector $(1, \omega^{\ell}, \ldots , \omega^{(N-1)\ell})^t$ for $ 0 \leq \ell \leq \ N-1$, and $D^{-1} R^N D$ is conjugate to the diagonal matrix ${\rm dia} [\lambda_0^N , \cdots , \lambda_{N-1}^N]$. Since $R^N$ and $D^{-1} R^N D$ have the same characteristic polynomials, the expression of $A_k(s)$ follows immediately. 
$\Box$  \par \vspace{.1in} \noindent
{\bf Remark.} In the $\lambda_\ell$-expression, $s^{1/N}$ is the variable $t$. Under the transformation $t \mapsto \omega t$, $\lambda_\ell$ is changed to $\lambda_{\ell -1}$ with $\lambda_{-1} := \lambda_{N-1}$. Therefore the $\lambda_\ell$-expressions in the above proposition do give the $s$-polynomials, which are equal to $A_k(s)$.
$\Box$

\section{Differential Equation of
BAMP Polynomials}
In this section, we discuss the differential
equations satisfied by BAMP polynomials
$P^{(L)}_Q$. For each $L$, we denote by 
${\cal P}^{(L)}(s)$ the $N$-vector with $P^{(L)}_Q$ as the
$j$th component for  $0 \leq j:=
a_{L, Q} \leq N-1$, i.e., ${\cal P}^{(L)}(s)$ is the $N$-vector, 
\be
{\cal P}^{(L)}(s) = \bigg(  
       P^{(L)}_{\underline{-L}}(s), 
       P^{(L)}_{\underline{-L-1}}(s),
      \ldots ,
 P^{(L)}_{\underline{-L-N+1}}(s) \bigg)^t \ \ .
\ele(vecP)
Then with the matrix $R$ in (\req(R)), we have
\be
(\sum_{j=0}^{N-1}t^j)^L = (1, t, \cdots, t^{N-1}) {\cal P}^{(L)}(s) \ , \ \ \ \ \ {\cal P}^{(L+1)}(s) = R {\cal P}^{(L)}(s) . 
\ele(PL)  
By differentiating $t\frac{d}{dt}$ the first relation of
(\req(PL)), then multiplying it by
$\sum_{j=0}^{N-1}t^j$, we obtain
\bea(ll)
&(1, t, \cdots, t^{N-1})
\cdot Ns(s-1){\cal P}^{(L) \prime} (s)  \\
=& 
(1, t,  \cdots, t^{N-1})
\cdot \bigg( L (t-1)  (\sum_{j=0}^{N-1}
jt^j)   - (s-1) {\rm
dia}[0, 1, \cdots, (N-1)] \bigg) {\cal P}^{(L)}(s) ,
\elea(diffP)
where we use the superscript prime $^\prime$ to denote the differentiation with respect to the variable $s$ for the rest of this section.  
By
\begin{eqnarray*}
(t-1)(\sum_{j=0}^{N-1}jt^j) = 
(N-1)s - \sum_{j=1}^{N-1}  t^j \ , & \\
(1, t, \cdots, t^{N-1}) t^j =
(1, t, \cdots, t^{N-1}) B_j
\ , & B_j : = \left( \begin{array}{cccc}
        0 & s I_j \\
        I_{N-j} & 0   
\end{array} \right) \ \ ,  \ 1 \leq j \leq N-1  \ ,
\end{eqnarray*}
the relation (\req(diffP)) is equivalent to the following
differential equation of ${\cal P}^{(L)}(s)$, 
\bea(l)
N s (s-1) {\cal P}^{(L) \prime}  = {\cal B} {\cal P}^{(L)}
\elea(eqnvP)
where ${\cal B}= L(N-1)s-L\sum_{j=1}^{N-1}B_j- 
(s-1) {\rm dia}[0, 1, \cdots, (N-1)] $, which has the following expression: 
$$
{\cal B}   =  
\left( \begin{array}{cccc}
      \delta_0  & -Ls & \cdots   & -Ls
\\
        -L & \delta_1 & \ddots &\vdots
\\ 
      \vdots  & \ddots  & \ddots &-Ls
\\ 
-L  & \cdots   & - L& \delta_{N-1} \\  
\end{array} \right) \ , \ \ \ \delta_j := L(N-1)s
-j(s-1) \ .
$$ 
By successive differentiations of $\frac{d}{ds}$ on (\req(eqnvP)), 
one can express $N^k s^k (s-1)^k \frac{d^k  {\cal P}^{(L)}}{ds^k}$ in terms 
of ${\cal P}^{(L)}$. In particular, for $k=2, 3$, we have 
\bea(ll)
N^2 s^2 (s-1)^2  {\cal P}^{(L) \prime \prime }  = & \bigg(  {\cal B}^2-N(2s-1) {\cal B} + N s(s-1){\cal B}^\prime \bigg) {\cal P}^{(L)} \\
N^3 s^3(s-1)^3  {\cal P}^{(L) \prime \prime \prime} =& \bigg(  {\cal B}^3-3N(2s-1) {\cal B}^2 +2 N^2 (3s^3-3s+1) {\cal B} \\& 
+ N s(s-1)(2{\cal B}^\prime {\cal B}+{\cal B}{\cal B}^\prime )- 2 N^2 (2s-1) s(s-1){\cal B}^\prime \bigg) {\cal P}^{(L)} \ .
\elea(k23d)
Using the relations between higher-order derivatives of ${\cal P}^{(L)}$, the following statement is expected to be true.

{\bf Conjecture 1}.  There exist diagonal matrices $D_k  
\ (0 \leq k \leq N-1) $ with entries (depending on $L$) in $\CZ[s]$, such that 
\be
N^N s^{N-1}(s-1)^{N-1} \frac{d^N  {\cal P}^{(L)}}{ds^N} + \sum_{k = 1 }^{N-1} N^k s^{k-1}(s-1)^{k-1} D_k \frac{d^k  {\cal P}^{(L)}}{ds^k} + D_0 {\cal P}^{(L)} = 0 \ .
\ele(conj1)
By which $P_Q^{(L)}$ satisfies a
$N$th order differential equation with the regular singular points at $s=0, 1$, and the expression  depends only on
$L + Q \pmod{N}$. $\Box$ \par 

We are going to demonstrate the solution of (\req(conj1)) for $N=2,3$. For $N=2$, by (\req(eqnvP)) and (\req(k23d)), one has
$$
\begin{array}{ll}
2 (s-1) s P^{(L) \prime}   =&   \left( \begin{array}{cc}
      Ls   & -Ls \\
        -L  & Ls-(s-1)  
\end{array} \right) P^{(L)} \ , \\
4 (s-1)^2 s^2 P^{(L) \prime \prime}= &  
\left( \begin{array}{cc}
      s(s+1)L^2 - 2s^2 L   &
-2s^2L^2 + s(3s-1)L   \\
        -2sL^2 +(5s-3)L & s(s+1)L^2- 2s(2s-1)L
+ 3(s-1)^2  
\end{array} \right)
       P^{(L)} \ .
\end{array}
$$
Then one obtains the solution of (\req(conj1)) for $N=2$:
$$
4 s(s-1) {\cal P}^{(L) \prime \prime} + 2 D_1  {\cal P}^{(L)\prime } + D_0 {\cal P}^{(L)} = 0 ,
$$ 
with $D_1 = - {\rm dia} \ [f_0, f_1]$ , $D_0 = {\rm dia} \ [g_0, g_1]$ for  
$$
f_0 = (2L-3)s+1 , \ \  
f_1 =(2L-5)s+3 \ ; \ \ \ 
g_0 = L(L-1) , \ \  g_1 = (L-1)(L-2) \ .
$$
The differential
equation of $P^{(L)}_Q$ is given by the hypergeometric differential
equations,
$$
\begin{array}{ll}
L_{:{}^{\rm even}_{\rm odd} }:& 4s(s-1)P_Q^{(L) \prime \prime} -2
f_{{}^Q_{1-Q}} (s) P_Q^{(L) \prime} +
g_{{}^Q_{1-Q}} (s) P_Q^{(L)}= 0 ,
\end{array}
$$
with the polynomial solutions: $P^{(L)}_Q (s)\sim F ( \frac{Q+1-L}{2},
\frac{Q-L}{2}, Q+
\frac{1}{2}; s) $ for even $ L$, and $P^{(L)}_Q (s) \sim  F ( 1- \frac{Q+L}{2},
\frac{1}{2}-\frac{Q+L}{2}, \frac{3}{2}-Q; s)$ for odd $L$.

For $N=3$,  (\req(eqnvP)) and (\req(k23d)) become 
$$
\begin{array}{cl}
3s(s-1){\cal P}^{(L) \prime} = {\cal B} \
{\cal P}^{(L)} ; &
9s^2(s-1)^2 {\cal P}^{(L) \prime \prime}  \ = \bigg( 
{\cal B}^2 -3(2s-1) {\cal B} +3s(s-1){\cal B}^\prime \bigg)
\ {\cal P}^{(L)} ; \\
27s^3 (s-1)^3{\cal P}^{(L) \prime \prime \prime}
= & \bigg(  {\cal B}^3 -9(2s-1) {\cal B}^2 +
18(3s^2-3s+1) {\cal B} +3s(s-1)(2{\cal B}^\prime {\cal
B}+{\cal B}{\cal B}^\prime) \\ & -18(s-1)s(2s-1){\cal B}^\prime
\bigg) \ {\cal P}^{(L)} \ ,
\end{array}
$$
with 
$$
{\cal B}= \left( \begin{array}{ccc}
      2Ls  & -Ls  & -Ls
\\
        -L & 2Ls-(s-1) & -Ls
\\ 
        -L & -L&2Ls-2(s-1)\\  
\end{array} \right) \ , \ \  {\cal B}^\prime  = 
\left( \begin{array}{ccc}
      2L & -L  & -L
\\
        0 & 2L-1 & -L
\\ 
        0 & 0&2L-2\\  
\end{array} \right) \ .
$$
The solution of (\req(conj1)) for $N=3$:
$$
27 s^2(s-1)^2 {\cal P}^{(L) \prime \prime \prime} +  9 s(s-1) D_2 {\cal P}^{(L) \prime \prime} + 3 D_1 {\cal P}^{(L) \prime} + D_0 {\cal P}^{(L)} = 0 \ ,
$$
is given by $
D_2 = -3 {\rm dia}\ [f_0, f_1, f_2]$, $ D_1 = {\rm dia}\ [g_0, g_1, g_2]$, $ D_0 = - (L-1) {\rm dia}\ [h_0, h_1, h_2]$ with
$$
\begin{array}{ll}
f_0 = (2L-4)s+2 , & g_0 =
3L^2s(4s-1)-3Ls(10s-7)+2(s-1)(10s-1) ,  \\
& h_0 = L \bigg( L(8s+1)-4(s-1)  \bigg) , \\
f_1 = (2L-6)s+4 , & g_1 =
3L^2s(4s-1)-3Ls(18s-15)+2(s-1)(31s-10) , \\
& h_1 = (L-2) \bigg( L(8s+1)- 12(s-1)  \bigg) , \\
f_2 = (2L-5)s+3 , & g_2 =
3L^2s(4s-1)-3Ls(14s-11)+2(s-1)(19s-4) , \\
& h_2 = L^2 (8s+1)-L(16s-7) + 6(s-1)  .
\end{array}
$$
Then one obtains the
third-order differential equation for
$P^{(L)}_Q$ for $L \equiv 0 \pmod{3}$: 
$$
27 s^2 (s-1)^2 P^{(L) \prime \prime
\prime}_Q - 27 s (s-1) f_Q P_Q^{(L) \prime
\prime}+3 g_Q P_Q^{(L) \prime} -(L-1)h_Q
P_Q^{(L)} = 0 \ , 
$$
The above differential equation
holds also for $P^{(L)}_{Q'}$ for a general $L$ with
$L+Q' \equiv Q \pmod{3}$.

\section{ Chebyshev-type Polynomials
${\bf \Pi^{(L)}_Q(c)}$} 
For the eigenvalue problem of (\req(SCPM)), one 
employs the theory of 
Onsager's algebra representation through 
the zeros $c_j$'s of some polynomials in the variable $c$, which relates to the variable $s$ in the previous two sections by the transformation,
$c=\frac{1+s}{1-s}$, equivalently, $s=
\frac{c-1}{c+1}$. The change of variables provides the one-to-one correspondence 
between $s \in (-\infty, 0)$
and $c \in (-1, 1)$. Accordingly, for the description of eigenvalues of   
(\req(SCPM)), one may convert a $s$-polynomial $F(s)$ to a $c$-polynomial $\Pi (c)$
via the relation 
$$
\Pi (c) : = (c+1)^{{\rm deg} \ F} F (
\frac{c-1}{c+1}) \ .
$$  
By which the
$c$-polynomial corresponding to the BAMP polynomial
$P^{(L)}_Q (s)$ will be denoted by
\be
\Pi^{(L)}_Q (c) = (c+1)^{b_{L,Q}}P^{(L)}_Q
(\frac{c-1}{c+1}) \ .
\ele(PiP)
Then the eigenvalue of (\req(SCPM)) (of size $L$) for the lowest state eigenvector in the $Q$-sector is given by $
E = 2Q k'+(N b_{L,Q} -(N-1)L) (1+k') - N \sum_{j=1}^{b_{L,Q}} \mp \sqrt{1 + 2k'c_j +k'^2 }$, 
where $c_j$ are zeros of $\Pi^{(L)}_Q (c)$  \cite{B}\footnote{The "$\cos \theta_j$" in formula (31) of \cite{B} is equal to "$-c_j$" here due to the convention, $c= \frac{1+s}{1-s}$, used in this aricle, also see formula (33) of \cite{R91}. }.
We shall call $\Pi^{(L)}_Q (c)$ the (generalized) Chebyshev-type polynomials for the reason being clear later on in the discussion of the case $N=2$.  

The reciprocal
relation of $s$-polynomials $P^{(L)}_Q(s)$ is
translated to the $*$-relation: $ \Pi^{(L)}_Q (c)
\rightarrow 
 \Pi^{(L) *}_Q(c) $. Here $p^*(c)$ is  the polynomial defined by $p^*(c) : = (-1)^{{\rm deg} \ p} p(-c)$ for  $p(c) \in \CZ[c]$.
Hence a $*$-symmetric polynomial $p(c)$ ( i.e., $p(c)= p^*(c)$) simply means an even or odd function according to the parity of the degree of $p(c)$. 
Then the relation , $
(pq)^*= p^* q^*$, holds. 
However, the equality, $( p + q)^*= p^*+ q^*$, is valid 
only for $p,  q$ with the same degree.
\begin{lemma} \label{lem:reciPi} $\Pi^{(L)}_Q(c)$ is a polynomial of degree 
$b_{L,Q}$ with the leading coefficient equal to $N^{L-1}$. And  
$\Pi^{(L) *}_Q(c)= \Pi^{(L)}_{Q'}(c)$  if and
only if 
$L+Q+Q' \equiv 0 \pmod{N} $.
\end{lemma}
{\it Proof.} By the positivity of all coefficients of  $P^{(L)}_Q(s)$, $\Pi^{(L)}_Q (c)$ is a $c$-polynomial of degree the
same  as the $s$-polynomial $P^{(L)}_Q (s)$, which is equal to $b_{L,Q}$. (Hence all the roots of $\Pi^{(L)}$  are real, confined in the open interval $(-1 , 1)$.) The leading coefficient of $\Pi^{(L)}_Q (c)$ is given by   $P^{(L)}_Q (1)= N^{L-1}$.
The $*$-symmetry relation of $\Pi^{(L)}_Q$s follows from Theorem \ref{thm:BAMP} $(i)$.
$\Box$ \par \vspace{.1in} \noindent
As a corollary of Lemma \ref{lem:reciPi}, one obtains the following results:
\begin{lemma} \label{lem:symPi} 
(i) When $N$ is odd, for each $L$ there is
exactly one $*$-symmetric $
\Pi^{(L)}_Q$ with 
$Q \equiv [\frac{N}{2}]L \pmod{N}$. 

(ii) When $N$ is even, $
\Pi^{(L)}_Q= \Pi^{(L) *}_Q $ holds only when $L$ is even, 
in which case there are exactly two $*$-symmetric
$\Pi^{(L)}_Q$ with $Q$ given by the relation $Q +\frac{L}{2}
\equiv 0 \pmod{\frac{N}{2}}$.
\end{lemma}
$\Box$ \par \vspace{.1in} \noindent
Denote the $c$-polynomial associated to $A_k(s)$ in Theorem \ref{thm:BAMP} $(ii)$ by 
\be
\alpha_k (c):= (c+1)^{k(N-1)}
A_k(\frac{c-1}{c+1}) 
, \ \ \ 1 \leq k \leq N \ .
\ele(ac)
By $b_{L+kN, Q}= b_{L, Q}+k(N-1)$, the  
substitution  $s= \frac{c-1}{c+1}$ in
(\req(recP)), together with the multiplication of $(c+1)^{b_{L+N^2,
Q}}$, gives rise to the recurrence
relation of
$\Pi^{(L)}_Q(c)$ for a fixed $Q$. Indeed by Theorem \ref{thm:BAMP} $(ii)$ and $(iii)$, we have the following result:
\begin{theorem} \label{thm:Pirec} 
With the polynomials
$\alpha_k(c)$ in $(\req(ac))$, we have the following
recurrence relation of $\Pi^{(L)}_Q$s for the same $Q$,
\be
\Pi^{(L+N^2)}_Q(c) + \ldots
+ \alpha_k(c) \Pi^{(L+(N-k)N)}_Q (c) + \ldots +
\alpha_N(c) \Pi^{(L)}_Q(c) = 0 \ .
\ele(recPi)
The polynomials  $\alpha_k(c)$ are $*$-symmetric, and $\alpha_1(c) = -N\Pi^{(N)}_0(c)$, 
$\alpha_N(c)=(-1)^N 2^{N(N-1)}$, $\alpha_k( \pm 1) = (-1)^k (\pm 2)^{k(N-1)}{N\choose k}$. For $k \geq 2$, the degree of $ \alpha_k(c)$ is less than $k(N-1)$ with $k(N-1)-{\rm deg} \alpha_k$ being a positive even integer. 
\end{theorem}
$\Box$ \par \vspace{.1in} \noindent
For $N=2,3, 4$, the recurrence relation (\req(recPi))   
takes the following explicit forms: 
\bea(ll)
N=2,  & \Pi^{(L+4)}_Q - 4c \Pi^{(L+2)}_Q + 4
\Pi^{(L)}_Q= 0 \\
N=3, & \Pi^{(L+9)}_Q -3(9c^2-5)\Pi_Q^{(L+6)} +48
\Pi^{(L+3)}_Q -64 \Pi^{(L)}_Q = 0 \ , \\
N=4, & \Pi^{(L+16)}_Q  
-32c(8c^2-7)\Pi_Q^{(L+12)} -128
(14c^2-17)\Pi^{(L+8)}_Q 
-2048c
\Pi^{(L+4)}_Q  \\
&
+4096
 \Pi^{(L)}_Q = 0 
\elea(Pi23)

In above discussions, the recurrence formula (\req(recPi)) follows from that of BAMP polynomials. However, it is more convenient to derive (\req(recPi)) directly from relations between $\Pi^{(L)}_{Q'}(c)$'s as follows. 
Parallel to (\req(vecP)), we define the following $N$-vector $\Pi^{(L)}$ and $N \times N$ matrix $\Re$:
$$
\Pi^{(L)}  =\bigg( 
       \Pi^{(L)}_{\underline{-L}}(c),
       \Pi^{(L)}_{\underline{-L-1}}(c),
      \ldots , 
 \Pi^{(L)}_{\underline{-L-N+1}}(c) 
\bigg)^t \ , \ \ \ \
\Re = \left( \begin{array}{ccccc}
       c+1  &  c-1 & \cdots   &  c-1 \\
        c+1 &  c+1 &  \ddots &\vdots \\ 
      \vdots  & \ddots & \ddots & c-1
\\ 
 c+1 & \cdots  &  c+1&  c+1 \\  
\end{array} \right)
$$ 
Denote
\be
{\cal C}^{(L)}=  \Delta^{(L+1)}
\Re \Delta^{(L) -1}(c+1)^{-1}
\ele(CL)
where $\Delta^{(L)}:= {\rm dia} \ [(
c+1)^{b_{L, \underline{- L}}}, (c+1)^{b_{L, \underline{-L-1}}}, \ldots , (c+1)^{b_{L,
\underline{-L-N+1}}}]$. 
For $0 \leq i , j \leq N-1$, we have $ b_{L+1, \underline{-L-1-i}} \geq b_{L, \underline{-L-j}}$, and the equality holds only for certain $(i, j)$'s with  $i \geq j$. Hence the entries of ${\cal
C}^{(L)}$ are all $c$-polynomials. Furthermore by  $b_{L+N, Q} = b_{L, Q}+(N-1)$, one has  
$\Delta^{(L+N)} = (c+1)^{N-1} \Delta^{(L)}$. This implies ${\cal
C}^{(L)}= {\cal C}^{(L+N)}$, equivalently,  $ {\cal C}^{(L)} = {\cal
C}^{(\underline{L})}$.
By (\req(PL)), one has the following
relation, 
\be
   \Pi^{(L+1)} = {\cal C}^{(L)}
\Pi^{(L)} \ ( = {\cal C}^{(\underline{L})}
\Pi^{(L)})  ,
\ele(RePi)
hence $
 \Pi^{(L+N)} = {\cal C}^{(L+N-1)} \cdots
{\cal C}^{(L+1)}{\cal C}^{(L)}
\Pi^{(L)} = (c+1)^{-N} \Delta^{(L+N)} \Re^N
\Delta^{(L) -1} \Pi^{(L)}$. 
Then we obtain 
\be
 \Pi^{(L+N)} = (c+1)^{-1} \Delta^{(L)} \Re^N
\Delta^{(L) -1} \Pi^{(L)}
\ .
\ele(RePiN)
Note that by $b_{Nk, Q}= (N-1)k$ for all $Q$, $ \Pi^{(L+N)} = (c+1)^{-1} \Re^N \Pi^{(L)}$ for $L \equiv 1 \pmod{N}$. 
The characteristic 
polynomials of the linear map in the right hand side of (\req(RePiN)) is given by
${\rm det}(x - (c+1)^{-1} \Re^N) =x^N + \sum_{k=1}^N \alpha_k(c) x^{N-k} $, which provides the coefficients in the recurrence relation (\req(recPi)) of $\Pi^{(L)}_Q$'s.

We now consider the case
$N=2$. By Lemma \ref{lem:reciPi}, $
\Pi^{(2k) *}_Q = \Pi^{(2k)}_Q$, $  
\Pi^{(2k+1) *}_0 = \Pi^{(2k+1)}_1 $ and 
 ${\rm deg} \Pi^{(2k)}_0 = {\rm
deg} \Pi^{(2k)}_1+1= {\rm
deg} \Pi^{(2k+1)}_Q= k $. We have  
$$
\Pi^{(2k)} =  \bigg( 
       \Pi^{(2k)}_0(c), 
  \Pi^{(2k)}_1(c) \bigg)^t \ , \ \ \ 
\Pi^{(2k+1)} =
\bigg( \Pi^{(2k+1)}_1(c) , 
  \Pi^{(2k+1)}_0(c) \bigg)^t \ , 
$$
and ${\cal C}^{(L)}$'s in (\req(CL)) are described by  $
{\cal C}^{(0)}= \left( \begin{array}{cc}
      1 &c-1 \\
       1 &c+ 1 
\end{array} \right) $, $ 
{\cal C}^{(1)} = \left( \begin{array}{cc}
          c+1& c- 1  \\
        1 & 1 
\end{array} \right)$. 
Then one has
$$
\Pi^{(2k+1)} = {\cal C}^{(0)}
\Pi^{(2k)} \ , \ \ \Pi^{(2k+2)}= 
{\cal C}^{(1)}
\Pi^{(2k+1)} \ ; \ \ \ \  
\Pi^{(2k+2)} = {\cal C}^{(1)}{\cal
C}^{(0)} \Pi^{(2k)} \ , \ \ \ \Pi^{(2k+3)} = {\cal C}^{(0)}{\cal
C}^{(1)} \Pi^{(2k+1)} \  .
$$
The matrices, ${\cal C}^{(1)}{\cal
C}^{(0)} =2 \left( \begin{array}{cc}
      c & 1 \\
       c^2-1& c 
\end{array} \right)$ and $ 
 {\cal C}^{(0)}{\cal
C}^{(1)} =2 \left( \begin{array}{cc}
      c & c+1 \\
       c-1& c 
\end{array} \right)$,
both satisfy the equation, $
x^2 -4c x + 4 = 0 $. 
Hence one has the following recurrence relation of $\Pi^{(L)}_Q$'s,
$$
\Pi^{(L+4)}_Q - 4c \Pi^{(L+2)}_Q + 4
\Pi^{(L)}_Q= 0 \ , 
$$
which is the same as in (\req(Pi23)) for $N=2$.
By  normalizing $\Pi^{(L)}_Q$ by a 2-power factor, one obtains the Chebyshev-relation: 
\begin{proposition} \label{prop:Che} 
For $\epsilon =0,1$, denote by
$F_k:= 2^{-k}\Pi^{(2k+\epsilon)}_Q $ for
$ k \in
\ZZ_{\geq 0}$. The  polynomials $F_k(c)$
satisfy the Chebyshev recurrence relations: $
F_{k+1} - 2c F_k + F_{k-1} = 0 $, 
with $F_0,
F_1$ given by
$$
\begin{array}{|l | l || c | c  |}
\hline
 \epsilon & Q   & F_0 &
F_1  \\
\hline
 0 & 0 & 1 &  c  \\
 0 & 1 & 0 & 1 \\  
 1 & 0 & 1 & 2c+1  
\\ 
 1 & 1 & 1 & 2c -1  \\
\hline
\end{array}
$$
As a consequence, $T_k= 2^{-k}\Pi^{(2k)}_0$, 
$ U_k= 2^{-k} \Pi^{(2k)}_1$ are the
Chebyshev polynomials of first and second
kind respectively\footnote{Here we use the
standard conventions: $T_k(x) = \cos (k {\rm
arccos} x)$; $U_k(x) = \sin (k {\rm
arccos} x)/ \sqrt{1-x^2} $.},  and \
$ 2^{-k}\Pi^{(2k+1)}_0 = U_{k+1}+U_k $,
$2^{-k}\Pi^{(2k+1)}_1 = U_{k+1}-U_k $.
\end{proposition}
$\Box$ \par \vspace{.1in} \noindent
It is known that Chebyshev polynomials form
a system of orthogonal polynomials  satisfying
a second-order differential
equation. Indeed by using the 
relations between
$\Pi^{2k+2}, \Pi^{2k+1}$ and
$\Pi^{2k}$, one obtains $
\Pi^{(2k+2)}_0 =  2c
\Pi^{(2k)}_0 + 2 (c^2-1) \Pi^{(2k)}_1 $ and $
\Pi^{(2k+2)}_1 =  2 
\Pi^{(2k)}_0  + 2 c 
\Pi^{(2k)}_1$. 
Then by induction argument, one can show 
$$
\frac{d\Pi^{(2k)}_0}{dc} = k \Pi^{(2k)}_1 , \ \ \
\  (1-c^2) \frac{d\Pi^{(2k)}_1}{dc} = -k
\Pi^{(2k)}_0 + c \Pi^{(2k)}_1 \ ,
$$
which are equivalent to $
\sqrt{1-c^2}\frac{d}{dc}\Pi^{(2k)}_0 = k \sqrt{1-c^2}
\Pi^{(2k)}_1 $, $\sqrt{1-c^2} \frac{d}{dc}
(\sqrt{1-c^2} \Pi^{(2k)}_1 ) = -k
\Pi^{(2k)}_0 $. 
Both $\Pi^{(2k)}_0$ and $ \sqrt{1-c^2}\Pi^{(2k)}_1$
are solutions of the equation, $
(\sqrt{1-c^2}\frac{d}{dc})^2 f = k^2 f $; then follow  
the differential
equations of $\Pi^{2k}_Q$,
$$
(1-c^2) \frac{d^2 \Pi^{(2k)}_0}{dc^2} -c 
\frac{d \Pi^{(2k)}_0}{dc} + k^2 \Pi^{(2k)}_0 = 0 \
, \ \ \ \ \
(1-c^2) \frac{d^2 \Pi^{(2k)}_1}{dc^2} -3c 
\frac{d \Pi^{(2k)}_1}{dc} + (k^2-1) \Pi^{(2k)}_1
 = 0 \ .
$$
By which,  $\Pi^{(2k)}_0(c)$ for $k \in \NZ$ form a system of orthogonal polynomials with weight $(1-c^2)^{\frac{-1}{2}}$; and the same  for 
$\Pi^{(2k)}_1(c)$'s.

\section{Differential Equation of 
Chebyshev-type Polynomials}
Along the path of discussions in the case $N=2$, one tends to find the
differential equations of Chebyshev-type polynomials $\Pi^{(L)}_Q$ for $N \geq 3$, especially for those $*$-symmetric ones described in Lemma \ref{lem:symPi}. 
By differentiating (\req(PiP)), one obtains the relation: 
$$
(c^2-1) \frac{d \Pi^{(L)}_Q}{dc} = b_{L,Q}(c-1)
\Pi^{(L)}_Q - (c+1)^{1+b_{L,Q} } s(s-1)
\frac{d P^{(L)}_Q }{ds} . $$
Using (\req(eqnvP)), one reaches the system of differential
equations of $\Pi^{(L)}_Q$'s:
\be
(c^2-1) \frac{d}{dc} \Pi^{(L)}  =  {\cal D} \Pi^{(L)}  \ ,
\ele(dPi)
where ${\cal D}= (d_{ij})_{0 \leq i, j \leq N-1}$ is the $N \times N$ matrix  with entries given by
$$
d_{ij} = \left\{ \begin{array}{ll}
-(c-1)\bigg( \frac{L(N-1)}{N}  -  b_{L,
\underline{-L-i}}
\bigg) - \frac{2i}{N}  & {\rm for}
\ i=j , \\
\frac{L}{N} (c+1)^{1+ b_{L, \underline{-L-i}}-b_{L,
\underline{-L-j} }} & {\rm for}
\  i> j ,
\\
\frac{L}{N} (c-1) (c+1)^{ b_{L, \underline{-L-i}}-b_{L,
\underline{-L-j} }} & {\rm for} \ i < j .
\end{array} \right.
$$
Note that one has
$$
b_{L, \underline{-L-i}}-b_{L,
\underline{-L-j}} = \left\{
\begin{array}{ll}
0 , & \underline{1-L} \leq j \ ,
i  , \ {\rm or} \ \ j, i < \underline{1-L} \ , \\
1 , & i <  \underline{1-L} \leq j \ , \\
-1 , & j < \underline{1-L} \leq i \ ,
\end{array}
\right.
$$
which imply entries of ${\cal D}$ are all $c$-polynomials. 
In particular, (\req(dPi)) for $N=3$ becomes 
\begin{eqnarray*}
(c^2-1) \frac{d}{dc} \left( \begin{array}{c}
       \Pi^{(3k)}_{0}(c)\\
       \Pi^{(3k)}_{2}(c)\\
 \Pi^{(3k)}_{1}(c) 
\end{array} \right)  & =& k \left( \begin{array}{ccc}
      0  &c^2-1 & c^2-1
\\
       1 & \frac{1-3c}{3k} & c-1\\ 
 1& c+1& \frac{-1-3c}{3k}
\end{array} \right)\left( \begin{array}{c}
       \Pi^{(3k)}_{0}(c)\\
       \Pi^{(3k)}_{2}(c)\\
 \Pi^{(3k)}_{1}(c) 
\end{array} \right) \ ; \\
(c^2-1) \frac{d}{dc} \left( \begin{array}{c}
       \Pi^{(3k+1)}_{2}(c)\\
       \Pi^{(3k+1)}_{1}(c)\\
 \Pi^{(3k+1)}_{0}(c) 
\end{array} \right)  &=& \frac{3k+1}{3} \left( \begin{array}{ccc}
      \frac{-2(c-1)}{3k+1}  &c-1 & c-1
\\
       c+1 & \frac{-2c}{3k+1}  &c-1\\ 
 c+1& c+1& \frac{-2(c+1)}{3k+1} 
\end{array} \right)\left( \begin{array}{c}
       \Pi^{(3k+1)}_{2}(c)\\
       \Pi^{(3k+1)}_{1}(c)\\
 \Pi^{(3k+1)}_{0}(c) 
\end{array} \right) \ ; \\
(c^2-1) \frac{d}{dc} \left( \begin{array}{c}
       \Pi^{(3k+2)}_{1}(c)\\
       \Pi^{(3k+2)}_{0}(c)\\
 \Pi^{(3k+2)}_{2}(c) 
\end{array} \right)  &=& \frac{3k+2}{3} \left( \begin{array}{ccc}
      \frac{-c+1}{3k+2}  &c-1 & c^2-1
\\
       c+1 & \frac{-4c+2}{3k+2}  &c^2-1\\ 
 1& 1& \frac{-c-3}{3k+2} 
\end{array} \right)\left( \begin{array}{c}
       \Pi^{(3k+2)}_{1}(c)\\
       \Pi^{(3k+2)}_{0}(c)\\
 \Pi^{(3k+2)}_{2}(c) 
\end{array} \right) \ .
\end{eqnarray*}
Due to the complicated expressions as indicated in above, it seems quite difficult to obtain a general form for the differential equations of  $\Pi^{(L)}_Q$ for an arbitrary $N$. However for  $L \equiv 0 \pmod{N}$,
the relation (\req(dPi)) becomes
$$
(c^2-1) \frac{d}{dc} \left( \begin{array}{c}
       \Pi^{(L)}_0 \\
       \Pi^{(L)}_{N-1} \\
    \vdots \\
    \vdots \\
\Pi^{(L)}_{1} \\ 
       \\  
\end{array} \right)    =  \frac{L}{N} 
\left( \begin{array}{ccccc}
0&c^2-1& \cdots  &\cdots & c^2-1
\\
1& \frac{-2-N(c-1)}{L}  &c-1 & 
\cdots& c-1 
\\
\vdots& c+1 & \frac{-4-N(c-1)}{L}  &\ddots &\vdots
\\ 
\vdots  & \vdots & \ddots & \ddots & c-1
\\ 
1&c+1 & \cdots & c+1 & 
\frac{-2(N-1)-N(c-1)}{L}
\end{array} \right) \left( \begin{array}{c}
       \Pi^{(L)}_{0}\\
      \Pi^{(L)}_{N-1} \\
\vdots \\
\vdots \\
 \Pi^{(L)}_{1} \\  
\end{array} \right)  \ .
$$
The first component in the above system gives the following simple relation:
\be
\frac{d\Pi^{(L)}_0}{dc} = \frac{L}{N}
\sum_{Q=1}^{N-1} \Pi^{(L)}_Q \ \ \ \ \ {\rm for} \ \ 
L \equiv 0 \pmod{N} \ .
\ele(dPi0)
By using (\req(dPi)), one expects the following statement to be true for all $N$ as in the case $N=2$:

{\bf Conjecture 2}. The polynomial $\Pi^{L}_Q(c)$ satisfies a $N$th-order differential equation with regular singular points at $c = \pm 1$. $\Box$ \par \vspace{.1in}

We are going to justify the above conjecture and derive the differential equations of $*$-symmetric polynomials $\Pi^{(L)}_Q$ (i.e., $\Pi^{(3k+Q)}_Q$) for $N=3$. For the rest of this section, we consider only the case $N=3$ where the degree $b_{L,Q}$ of $\Pi^{(L)}_Q(c)$ are 
  given by 
$$
2k  \ = b_{3k, 0} = b_{3k, 1}+1  = b_{3k, 2}+1 \ = 
b_{3k+1, 0} = b_{3k+1, 1} = b_{3k+1, 2} \ =
b_{3k+2, 0}-1 = b_{3k+2, 1}-1 = b_{3k+2, 2} .
$$
By (\req(Pi23)), one has  
the four-term recurrence relation of $\Pi^{(L)}_Q$s :  
\be
 \Pi^{(3(k+3)+\epsilon)}_Q
-3(9c^2-5)\Pi_Q^{(3(k+2)+\epsilon)} +48
\Pi^{(3(k+1)+\epsilon)}_Q -64
\Pi^{(3k+\epsilon)}_Q = 0 \ ,
\ele(N3r)
with $\Pi^{(3k+\epsilon)}_Q$ for $k=0,1,2$ given by
$$
\begin{array}{|l | l || c | c |c |}
\hline
 \epsilon & Q   & \Pi^{(\epsilon)}_Q &
\Pi^{(3+\epsilon)}_Q &
\Pi^{(6+\epsilon)}_Q \\
\hline
 0 & 0 & 1 &  9c^2-5 & 243c^4
-270c^2+43   \\
 0 & {}^1_2 & 0 &  9c \pm 3& \
243c^3 \pm 81c^2-135c \mp 21 \\ 
 1 & {}^0_2 & 1 & 27c^2 \pm 18c-5 &
729c^4 \pm 486c^3-540c^2 \mp 270c+43  
\\ 
 1 & 1 & 1 & 27c^2-11 &
729c^4 - 702c^2 +85  
\\  
 2 & {}^0_1 & 3c \pm 1 &
81c^3 \pm 27c^2-57c \mp 11 & 2187c^5 \pm 729c^4-2754c^3
\mp 702c^2+711c \pm 85    \\
 2 & 2 &3 & 81c^2-21 &
2187c^4 - 1782c^2 +171  \\
\hline
\end{array}
$$
By Lemma \ref{lem:symPi} $(i)$, 
$$
\Pi^{(3k+Q) *}_Q  = \Pi^{(3k+Q)}_Q, \
\Pi^{(3k) *}_2  = \Pi^{(3k)}_1 , \ 
\Pi^{(3k+1) *}_2  = \Pi^{(3k+1)}_0 , \ 
\Pi^{(3k+2) *}_1  = \Pi^{(3k+2)}_0.
$$ 
Hence one can express the relation (\req(RePi)) in terms of
$\Pi^{(3k+Q)}_Q$ and $\Pi^{(3k)}_1 , \Pi^{(3k+1)}_0 ,
\Pi^{(3k+2)}_0 $:
\bea(l)
\Pi^{(3k+1)}_0 = \Pi^{(3k)}_0 +
(c+1)(\Pi^{(3k)}_1 +  \Pi^{(3k) *}_1), \ \ \\ 
\Pi^{(3k+1)}_1 =  \Pi^{(3k)}_0 +
(c-1)\Pi^{(3k)}_1 +(c+1)
\Pi^{(3k) *}_1 , \\
\Pi^{(3k+2)}_0 =(c-1) \Pi^{(3k+1)}_0 +
(c+1)(  \Pi^{ (3k+1) *}_0  + 
 \Pi^{(3k+1)}_1 ) , \\  
\Pi^{(3k+2)}_2 = 
\Pi^{(3k+1)}_0 + \Pi^{(3k+1)}_1 +
\Pi^{(3k+1) *}_0 \ , \\
\Pi^{(3k+3)}_0 = (c-1) \Pi^{(3k+2)}_0 +
(c+1) \Pi^{(3k+2) *}_0 + (c^2-1) 
\Pi^{ (3k+2)}_2 ,  \\
\Pi^{(3k+3)}_1 =  \Pi^{(3k+2)}_0 +
\Pi^{(3k+2) *}_0 + (c+1)  \Pi^{( 3k+2)}_2 \ .
 \elea(REc3)
By (\req(RePiN)) with $(N, L)=(3, 3k)$, one has
$$
 \left( \begin{array}{c}
\Pi^{(3k+3)}_0 \\
\Pi^{(3k+3)}_1\\
\Pi^{(3k+3)}_2
\end{array}\right)
= \left(
\begin{array}{ccc}
   9c^2-5&   3(c^2-1)(3c-1) &3(c^2-1)(3c+1) \\
3(3c+1) &9c^2-5 & 3(c+1)(3c-1)\\
3(3c-1) &3(c-1)(3c-1)&9c^2-5\\ 
\end{array} \right) \left( \begin{array}{c}
\Pi^{(3k)}_0 \\
\Pi^{(3k)}_1\\
\Pi^{(3k)}_2
\end{array}\right) 
\ .
$$
For convenience of notations, we denote  $P_k:=
\Pi_0^{(3k)}, H_k:= \Pi_1^{(3k)}$. One has
\bea(ll)
P_{k+1} &= (9c^2-5)P_k+3(c^2-1)\bigg( (3c-1)H_k
+(3c+1) H_k^* \bigg), \\
H_{k+1} &= 3(3c+1)P_k + (9c^2-5)H_k
+3(c+1)(3c-1)H_k^* , \\
H_{k+1}^*&= 3(3c-1)P_k+ 3(c-1)(3c+1)H_k+
(9c^2-5)H_k^* \ .
\elea(Pi3)
\begin{lemma}  \label{lem:diffE} 
Denote by $E_k= H_k-H_k^*$. Then the following
relations hold,
\bea(c)
k(H_k+H_k^*)  = \frac{d P_k}{dc} ,
 \ \ \ \ \ \ \
\frac{k(3k+1)}{3} E_k = -
 (c^2-1)\frac{d^2P_k}{dc^2}
+(k-1)c\frac{dP_k}{dc}+ 2k^2 P_k , \\
\frac{k(3k+1)}{3}\frac{dE_k}{dc}=
 (k+1)c\frac{d^2P_k}{dc^2} 
-((k^2-1)-\frac{8}{9(c^2-1)}
)\frac{dP_k}{dc}
-\frac{2k^2(k+1)c}{c^2-1}P_k \ .
\elea(eqn)
\end{lemma}
{\bf Proof.} The first relation in (\req(eqn)) 
follows from (\req(dPi0)) for $N=3$. 
We now show the other two relations by induction on $k$. The relations are easily verified for
$k=1$. Assume (\req(eqn)) holds for
$k$. By (\req(Pi3)) (\req(eqn)), we have
$$
\begin{array}{ll}
P_{k+1} &= (9c^2-5)P_k+ \frac{9}{k} c(c^2-1)
\frac{dP_k}{dc}-3(c^2-1)E_k ; \\
\frac{dP_{k+1}}{dc} & = (k+1)(18cP_k
+\frac{1}{k}(18c^2-8)\frac{dP_k}{dc}-6cE_k) ;  \\
E_{k+1} & = 6P_k+6c (H_k+H_k^*)-2E_k \ \ 
= 6P_k+ \frac{6}{k}c\frac{dP_k}{dc}-2E_k .
\end{array}
$$
By differentiating the 2nd and 3rd relations in above, one obtains:
$$
\begin{array}{ll}
\frac{c^2-1}{2(k+1)}\frac{d^2P_{k+1} }{dc^2}
&=  
(9(2k+1)c^2-8k-9) P_k
+\frac{1}{k}c(9(2k+1)c^2-16k-13)
\frac{dP_k}{dc} \\
& - 
 \frac{1}{3}( 9(2k+1)c^2-12k-13) E_k ; 
\\
(c^2-1)\frac{dE_{k+1}}{dc} &= - 12kc P_k
+\frac{4}{3}(-9c^2+
\frac{6k+4}{k} )
\frac{dP_k}{dc}
+4kc E_k \ .
\end{array}
$$
By which follow the
second and third identities of (\req(eqn))
for $k+1$. 
$\Box$ \par \vspace{0.1in} \noindent
\begin{theorem} \label{thm:Eq3} 
For $L \equiv Q \pmod{3}$, the polynomial $\Pi^{(L)}_Q$
satisfies the following $3$rd order differential
equation:
\begin{eqnarray}
27(c^2-1) \bigg( (c^2-1) \frac{d^3\Pi^{(L)}_Q}{dc^3}+
2(Q+2)c
\frac{d^2\Pi^{(L)}_Q}{dc^2}\bigg)
- \bigg( 9(L-2Q-2)(L+2Q+3)(c^2-1) \nonumber \\
-12(3Q^2+3Q+2) \bigg) 
\frac{d\Pi^{(L)}_Q}{dc}
-2(L-Q)(L+2Q)(L+2Q+3)c
\Pi^{(L)}_Q
= 0 \ . \label{diffN3}
\end{eqnarray}
\end{theorem}
{\it Proof.} In this proof, $P_k,
H_k, E_k$ denote the $c$-polynomials in
Lemma \ref{lem:diffE}, and the $c$-derivatives of
a function $f(c)$ 
 will be denoted by $f^\prime, f^{\prime \prime}, \cdots$
etc. Among the relations in (\req(eqn)), the differentiation of the second one is the same as the left hand side of the third one, by which one obtains the constraint of $P_k$, i.e. $\Pi^{(3k)}_0$, which is the differential equation  (\ref{diffN3}) for $Q=0$. 

By (\req(REc3)) and (\req(eqn)), we have
\bea(cl)
\Pi^{(3k+1)}_1&= P_k + c (H_k+H_k^*)-E_k \  = P_k
+ \frac{c}{k}P_k^{\prime } -E_k \\
(c^2-1)\Pi_1^{(3k+1) \prime } & = 2kcP_k + (2(c^2-1)- \frac{2}{3k})
P_k^{\prime} + \frac{2c}{3}E_k \ , 
\elea(1eq)
which implies
\bea(cl)
2 c \Pi^{(3k+1)}_1 + 3
(c^2-1)\Pi_1^{(3k+1) \prime } & = 2
(3k+1) ( c P_k  +   \frac{(c^2-1)}{k}  
P_k^{\prime } )  \ , \\
\frac{2}{3k}
P_k^{\prime } - \frac{2c}{3}E_k  &= 
-(c^2-1)\Pi_1^{(3k+1) \prime }
+2k( cP_k + \frac{(c^2-1)}{k}P_k^{\prime })
\\ &= -\frac{(c^2-1)}{3k+1}
\Pi_1^{(3k+1) \prime} +
\frac{2kc}{3k+1} \Pi^{(3k+1)}_1 \ .
\elea(Q1rel)
Differentiating the 2nd relation in (\req(1eq)),
then using the 2nd one in (\req(eqn)),
we have 
$$
\begin{array}{ll}
(c^2-1)\Pi_1^{(3k+1) \prime \prime} + 2c
\Pi^{(3k+1) \prime}_1 +(
\frac{2}{3k}
P_k^{\prime } - \frac{2c}{3}E_k )^{\prime}
& = 2kP_k  +2(k+2)c P_k^{\prime }+ 2(c^2-1)
P_k^{\prime \prime} \\
&=
2k(2k+1)P_k +2(2k+1)c  P_k^{\prime }-
\frac{2k(3k+1)}{ 3} E_k  \ .
\end{array}
$$
By the last equality of (\req(Q1rel)), we obtain
$$
2(2k+1)P_k +\frac{2(2k+1)c}{k}P_k^{\prime }-
\frac{2(3k+1)}{ 3} E_k 
= \frac{3(c^2-1)}{3k+1}
\Pi_1^{(3k+1) \prime \prime} +
\frac{8c}{3k+1}
\Pi_1^{(3k+1) \prime } +
\frac{2}{3k+1} \Pi^{(3k+1)}_1 \ .
$$
Then from the first relation of (\req(1eq)), one
arrives the following relation:
$$
\frac{2(3k+2)}{3}P_k
+\frac{2(3k+2)c}{3k}P_k^{\prime }
= \frac{3(c^2-1)}{3k+1}
\Pi_1^{(3k+1) \prime \prime} +
\frac{8c}{3k+1}
\Pi_1^{(3k+1) \prime } -
\frac{2(9k^2+6k-2)}{3(3k+1)} \Pi^{(3k+1)}_1 \ .
$$
By which and the first relation in
(\req(Q1rel)), one solves $P_k, P_k^\prime$
in terms of $\Pi^{3k+1}_1$, $\Pi^{3k+1 \prime}_1$ and $\Pi^{3k+1 \prime \prime}_1$:
$$
\begin{array}{rl}
 -2(3k+1)(3k+2) P_k  = &9(c^2-1)^2
\Pi_1^{(3k+1) \prime \prime}+
9(2-k)c(c^2-1) 
\Pi_1^{(3k+1) \prime } \\
&+ 2(9k^2+6k-2 - 9k(k+1)c^2) \Pi^{(3k+1)}_1 \ , 
   \\
\frac{2(3k+1)(3k+2)}{3k}P_k^{\prime }
=& 3c(c^2-1)
\Pi_1^{(3k+1) \prime \prime} +
3(2-k)c^2 +(3k+2)
\Pi_1^{(3k+1) \prime }  -
6k(k+1)c \Pi_1^{(3k+1)} .
\end{array} 
$$
By comparing the derivative of the first equality 
with the second relation in above, one arrives the differential equation (\ref{diffN3}) for $Q=1$, i.e. $\Pi^{(3k+1)}_1$.

By (\req(REc3)) and (\req(1eq)), we have
\bea(cl)
\Pi^{(3k+2)}_2 
&= 
3P_k+3c(H_k+H_k^*)-E_k = 
3P_k+\frac{3}{k}cP_k^{\prime}-E_k \ , \\
(c^2-1)\Pi_2^{(3k+2) \prime} &= 6kcP_k+(6c^2-\frac{12k+8}{3k}) P_k^\prime
-2kcE_k \ ,
\elea(2eq)
which implies
\be
 P_k^\prime
 =\frac{3k^2}{2(3k+2)}c\Pi^{(3k+2)}_2  -
\frac{3k}{4(3k+2)}(c^2-1)\Pi^{(3k+2)
\prime}_2
\ .
\ele(2eqPp)
By differentiating (\req(2eqPp)), and using the
second relation of (\req(eqn)), we obtain the
following identity from expressions of
$(c^2-1) P_k^{\prime \prime}$: 
$$
\begin{array}{ll}
2k^2 P_k 
+(k-1)c P_k^{\prime} - \frac{k(3k+1)}{3} E_k
 = & -
\frac{3k}{4(3k+2)}(c^2-1)^2\Pi^{(3k+2) \prime
\prime}_2 +
\frac{6k(k-1)}{4(3k+2)} c
(c^2-1)\Pi^{(3k+2) \prime}_2 
\\
&+\frac{3k^2}{2(3k+2)} (c^2-1)\Pi^{(3k+2)}_2 .
\end{array}
$$
Then by (\req(2eqPp)) and the first relation of (\req(2eq)),  we obtain 
$$
 12( k+1 )(3k+2) P_k =
9(c^2-1)^2
\Pi^{(L+2) \prime \prime}_2
+36c(c^2-1)\Pi^{(L+2) \prime}_2 +2
\bigg((3k+4)(6k+1)-9k(2k+3)c^2 \bigg)
\Pi^{(L+2)}_2 \ .
$$
By comparing (\req(2eqPp)) with the derivative of the above relation, one obtains the
differential equation (\ref{diffN3}) for $Q=2$.
$\Box$ \par \vspace{0.1in} 

The $*$-symmetric $\Pi^{(3k+Q)}_Q$ can be regarded as $N$=3
version of the Chebyshev polynomials appeared in the case $N$=2. The rest 
$\Pi^{(L)}_Q$  can be expressed in
terms of 
$\Pi^{(3k+Q)}_Q$'s. Indeed, by Lemma \ref{lem:symPi} $(i)$,  
$\Pi^{(L)}_Q$ for $L
\not\equiv Q \pmod{3}$ are determined by the following $*$-symmetric or antisymmetric polynomials: $ \Pi^{(3k)}_1-
\Pi^{(3k) *}_1, 
\Pi^{(3k+1)}_0+ 
\Pi_0^{(3k+1) *}, \Pi^{(3k+2)}_0  -
\Pi_0^{(3k+2) *}$, and  $ 
\Pi^{(3k)}_1+ \Pi_1^{(3k) *}, 
\Pi^{(3k+1)}_0- 
\Pi_0^{(3k+1) *} , \Pi^{(3k+2)}_0  +
\Pi_0^{(3k+2) *}$.
By (\req(REc3)) and Lemma \ref{lem:diffE}, we
have the relations,
$$
\begin{array}{llll}
\Pi^{(3k)}_1 + \Pi_1^{(3k) * } & =
\frac{1}{k}\frac{d\Pi^{(3k)}_0}{dc} , &  \Pi^{(3k)}_1 - \Pi_1^{(3k) *}&= 2 \Pi^{(
3k-1)}_2 ; \\
\Pi^{(3k+1)}_0 + \Pi_0^{(3k+1) *} &=
\Pi^{(3k+2)}_2 - \Pi^{(3k+1)}_1 , &
\Pi^{(3k+1)}_0 - \Pi_0^{(3k+1) *} &= 
\frac{2}{k}\frac{d\Pi^{(3k)}_0}{dc} ; \\
\Pi^{(3k+2)}_0 + \Pi_0^{(3k+2) *}&= 2c
\Pi^{(3k+2)}_2
-\frac{4}{k}\frac{d\Pi^{(3k)}_0}{dc} , &
\Pi^{(3k+2)}_0 - \Pi_0^{(3k+2) *}&= 2 
\Pi^{(3k+1)}_1 \ .
\end{array}
$$
By which, one can express the functions $\Pi^{(L)}_Q$ in terms of $\Pi^{(3k+Q )}_Q$'s.
\par \vspace{0.1in}\noindent
{\bf Remark.} 
As $\Pi^{(3k+Q)}_Q$ is an even $c$-polynomial 
of degree $2k$, hence depending only on $c^2$, one can express  $\Pi^{(3k+Q)}_Q(c)$ as a polynomial of the variable $\sigma = 1 -c^2$ : $ \Pi^{(3k+Q)}_Q (c) = \pi^{(k)}_Q (\sigma ) $  
for some $\sigma$-polynomial $\pi^{(k)}_Q$ of degree $k$, with all its roots  in the open interval $(0, 1)$. By (\req(N3r)), $\pi^{(k)}_Q$ satisfy the 4-terms recurrence relation: $
 \pi^{(k+3)}_Q -3(4- 9 \sigma)\pi_Q^{(k+2)} +48
\pi^{(k+1)}_Q -64
\pi^{(k)}_Q = 0$, 
with $\pi^{(k)}_Q$ for $k=0,1,2$ given by 
$$
\begin{array}{| l || c | c |c |}
\hline
  Q   & \pi^{(0)}_Q &
\pi^{(1)}_Q &
\pi^{(2)}_Q \\
\hline
  0 & 1 &  -9\sigma +4  & 243\sigma^2
-216 \sigma + 16   \\
  1 & 1 & -27 \sigma + 16 &
729\sigma^2 - 756 \sigma + 112 
\\  
  2 &3 & -81 \sigma +60 &
2187 \sigma^2 -2592 \sigma + 576  \\
\hline
\end{array}
$$
Using the relations, $
\frac{d}{dc} = -2c \frac{d}{d \sigma}$, $ 
\frac{d^2}{dc^2}= 4(1-\sigma)
\frac{d^2}{d\sigma^2}-2 \frac{d}{d\sigma}$ and $ 
\frac{d^3}{dc^3} = -4c \bigg(2(1-\sigma)
\frac{d^3}{d\sigma^3}-3 \frac{d^2}{d\sigma^2} \bigg)$, 
one can convert the equation (\ref{diffN3})  
into the following differential equation of $\pi^{(k)}_Q$:
$$
\begin{array}{l}
\frac{d}{d\sigma}\bigg( 4 \sigma^2  (1-\sigma)
\frac{d^2\pi^{(k)}_Q}{d\sigma^2} -2\sigma 
( (1+2Q)\sigma   -2Q)
\frac{d\pi^{(k)}_Q }{d\sigma} + 
( (3k^2+(2Q+1)k-Q^2+3Q-2)
\sigma \\ +\frac{4(3Q^2-6Q+2)}{9} )
\pi^{(k)}_Q \bigg) 
= -(k-2+Q)(k-1+Q)(k+1)
\pi^{(k)}_Q \ .
\end{array}
$$
$\Box$ \par 
\section*{ Acknowledgments} 
This paper is mainly based on the joint work with G.von Gehlen \cite{GRo}, except that a systematic account on mathematical structures of BAMP and Chebyshev-type polynomials for an arbitrary $N$ is presented, and certain parts contain  variations from the theme in \cite{GRo} by adding more rigorous mathematical arguments here. This is an occasion to thank G.von Gehlen for the stimulating collaboration and correspondences. This work has been supported in part by NSC 93-2115-M-001-013, Taiwan.

\end{document}